%% file: response.tex
\documentclass[11pt,reqno,letter]{amsart}
\usepackage{amsmath}

\usepackage{amsfonts}
\usepackage{amssymb}
\usepackage{amsthm}

\usepackage{graphicx}
\usepackage{hyperref}
\usepackage[left=1.25in,right=1.25in,top=1.5in,bottom=1.25in]{geometry}
\usepackage{multirow}
\usepackage{verbatim}
\usepackage{float}
\usepackage{rotating}
\usepackage[dvipsnames]{xcolor}
\usepackage[normalem]{ulem}
\usepackage{booktabs}

\usepackage{tabu}

\usepackage{float}
\usepackage{booktabs}

\usepackage{subcaption}

\theoremstyle{definition}

\newtheorem*{example*}{Example}

\usepackage{calc,tikz,nicefrac}

\usepackage[section]{placeins}

\usetikzlibrary{positioning}
\usepackage{mathtools}

\usepackage{bm}

\usepackage[authoryear]{natbib}
\usepackage{pdflscape}
\usepackage{afterpage}
\usepackage{capt-of}

\def\independenT#1#2{\mathrel{\setbox0\hbox{$#1#2$}%
    \copy0\kern-\wd0\mkern4mu\box0}}

\usepackage{neuralnetwork}

\begin{document}

\title[Response]
{A Response to Philippe Lemoine's Critique on our Paper ``Causal Impact of Masks, Policies, Behavior on Early Covid-19 Pandemic in the U.S.''}

\author{Victor Chernozhukov}
\address{Department of Economics and Center for Statistics and Data Science, MIT,  MA 02139}
\email{vchern@mit.edu}
\author{Hiroyuki Kasahara}
\address{ Vancouver School of Economics, UBC, 6000 Iona Drive, Vancouver, BC.}
\email{hkasahar@mail.ubc.ca}

\author{Paul Schrimpf}
\address{ Vancouver School of Economics, UBC, 6000 Iona Drive, Vancouver, BC.}
\email{schrimpf@mail.ubc.ca}

\date{\today}

\maketitle


\section{Context}

Our paper  ``Causal Impact of Masks, Policies, Behavior on Early Covid-19 Pandemic in the U.S''  \citep{chernozhukov2021} found mitigating effects of masks and personal behavior and could not rule out significant mitigation effects of various ``shut down” policies.  The paper conducted a battery of robustness checks reporting documenting the degree of robustness of various findings.   

Recently, Phillippe Lemoine posted a critique of our
paper  \citep{chernozhukov2021}   at his post titled \href{https://cspicenter.org/blog/waronscience/lockdowns-econometrics-and-the-art-of-putting-lipstick-on-a-pig/}{``Lockdowns, econometrics and the art of putting lipstick on a pig''}.
Although Lemoine's critique appears ideologically driven and overly emotional, some of his points are worth addressing. In particular,  the sensitivity of our estimation results for (i)  including ``masks in public spaces”  and (ii) updating the data seems important critiques and, therefore, we decided to analyze the updated data ourselves. This note summarizes our findings from re-examining the updated data and responds to Phillippe Lemoine's critique on these two important points. We also briefly discuss other points Lemoine raised in his post.

\section{Our Response}

\subsection{Summary}
After analyzing the updated data, we find evidence that reinforces the conclusions reached in the original study. 

\begin{enumerate}
\item Lemoine showed that replacing “masks for employees” (business mask mandates) by “masks in public spaces” (public mask mandates)  changes the effect estimate from negative to slightly positive.  This critique is  an obvious mistake because dropping “masks for employees”  variable introduces a confounder bias in estimating the effect of  “masks in public spaces.” When we include both “masks for employees only” and “masks in public spaces”  in the regression, the estimated coefficients of both variables are significantly negative in the original data.   Lemoine’s argument seems to be an obvious but honest mistake.\smallskip

\item  The second main point of Lemoine's critique is non-robustness for the data update. However,  Lemoine has not validated the new data. We find that  the timing of the first mask mandate for Hawaii is mis-coded in the updated data.  After correcting this data mistake,  the estimated coefficients of  “masks for employees only” and “masks in public spaces” are significantly negative. Similar results are obtained when we estimate by dropping Hawaii from the sample.  This critique is also an honest (though not obvious) mistake.\smallskip

\item   Lemoine analyzed the updated data that kept the original sample period from March 7 to June 3, 2020.  The negative effects of masks on case growth continue to hold when we extend the endpoint of the sample period to July 1, August 1, or September 1. With the extended data,  the estimated coefficients of “masks in public spaces” range from $-0.097$ to $-0.124$ with standard errors of $0.02\sim 0.03$ in Tables  \ref{tab:PtoY-6}-\ref{tab:PtoY-7}, and are roughly twice as large as those of “masks for employees only.” A preprint version of our paper was available in ArXiv in late May of 2020 and was submitted
for publication shortly after, which is why we did not analyze either the updated data or the extended data in our original paper. \smallskip

\item Lemoine applies the regression method to the simulated data from the SIR model and claims that ``the kind of model used by Chernozhukov et al. performs extremely poorly.” But, if we closely look at his results, it is difficult to conclude that our model performs ``extremely poorly'' because his regression estimates using the simulated data correctly indicate that non-pharmaceutical interventions reduce the case growth. The counterfactual effects implied by the estimated regression model are also within the ``confidence interval'' he computed from simulating the SIR models. Therefore, this critique does not make sense to us. \smallskip

\item  Lemoine claims that he did a placebo test and easily found spurious effects.  We implemented our placebo test by generating 500 samples, where each sample is identical to the original sample except that state-level mask variables are replaced with those of other states that are randomly selected through permutation. The result of our placebo test indicates that there won't be any systematic spurious effects we may find from our regression analysis. \smallskip

\item Lemoine also asserts that our MA(7) “smoothing” is “inappropriate” without any justification.  Using unsmoothed data can only makes sense if we artificially try to lower the precision of the statistical inference and it seems to us that this is what Lemoine tries to achieve (given his ideological stance). Therefore, this point does not make sense to us.\smallskip

\item Lemoine's critique also concerns the width of the confidence intervals. In a counterfactual simulation of the dynamic effect of business closures in our basic model, the 90\% confidence interval for confirmed deaths counts is from [-4\%, +64\%]. Lemoine' concludes therefore that business closures had zero effect because the null hypothesis of zero effect cannot be rejected at 10\% significance.  This is incorrect interpretation of confidence intervals.  Confidence intervals collect all point hypotheses that can not be rejected, in particular, in this case we can not rule out large or no effects on deaths of the mitigation policies (for case counts, the 90\% CIs based on baseline model was ruling out zero effects). 

\item When we use the extended sample period to August 1 of 2020, the coefficients of mask policy variables are more precisely estimated with the shorter confidence interval because many states implemented the  mask mandates  in June and July, helping identify the mask coefficients better. Lemoine could have analyzed the updated data with the extended sample period if he had a concern over the wide confidence interval. 
 
\end{enumerate}

\subsection{Detailed response}

\subsubsection{Including ``mandating face masks for everyone''}

In our paper, we analyze the effect of mask mandates for employees in
public-facing businesses. Our dataset also contains information on
mask mandates for all people in public spaces. Our decision to focus
on the former and not the latter is largely a historical
coindcidence. We used the excellent [“COVID-19 US state policy
database.” from  Raifman J, Nocka K, Jones D, Bor J, Lipson S, Jay J,
and Chan P. (2020).](www.tinyurl.com/statepolicies). This database has
been updated throughout the pandemic. When we began our paper, it
didn't include any information on mask mandates. When information on
mask mandates were added, and we began analyzing them, mask mandates for all people were quite rare, so we chose to focus on the more
common policy of mask mandates for businesses. However, by the time we
were finalizing the published version of our paper, public mask
mandates were common enough that we could have analyzed them
separately. Let's do that now.

\begin{figure}[!htbp]\caption{Portion of states with each
    policy for the updated data \label{fig:policyportion}}\medskip 
  \begin{minipage}{0.8\linewidth}
    \begin{tabular}{cc}
      \includegraphics[width=0.45\textwidth]{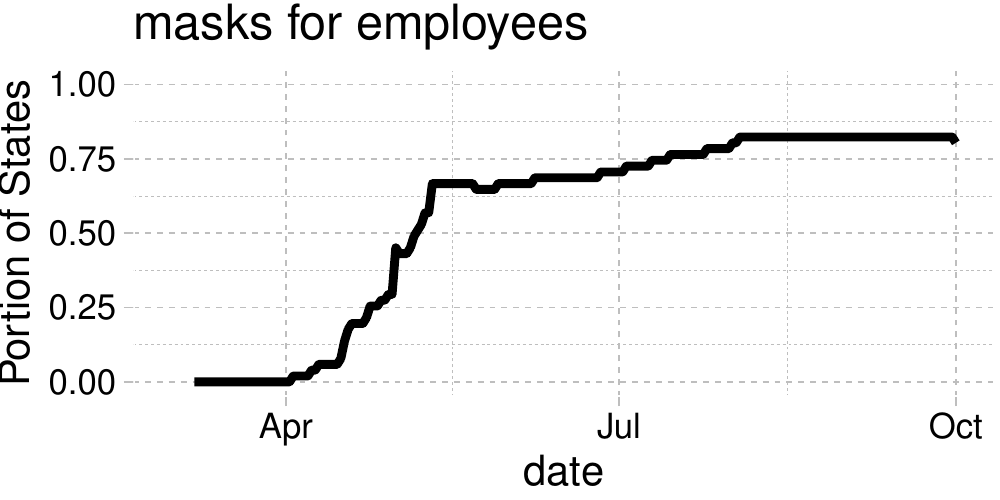}
      &
      \includegraphics[width=0.45\textwidth]{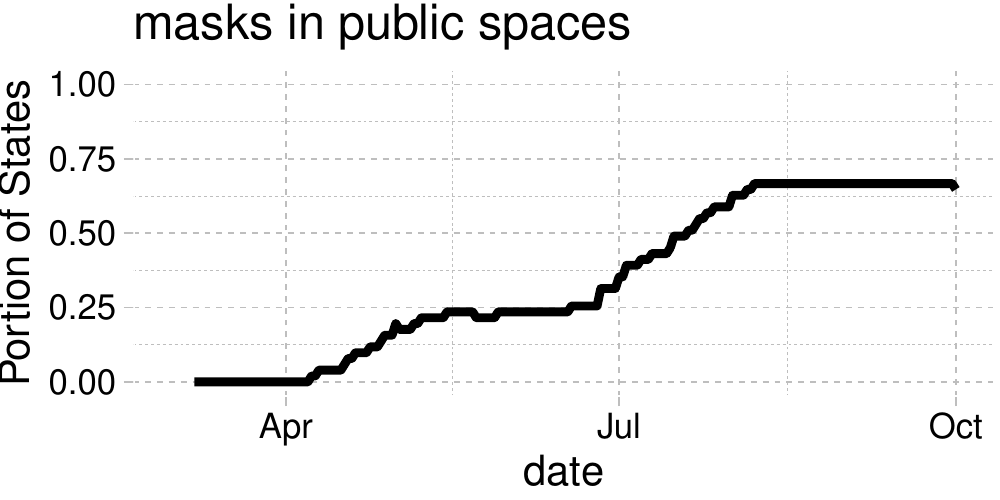}\\
  
        \includegraphics[width=0.45\textwidth]{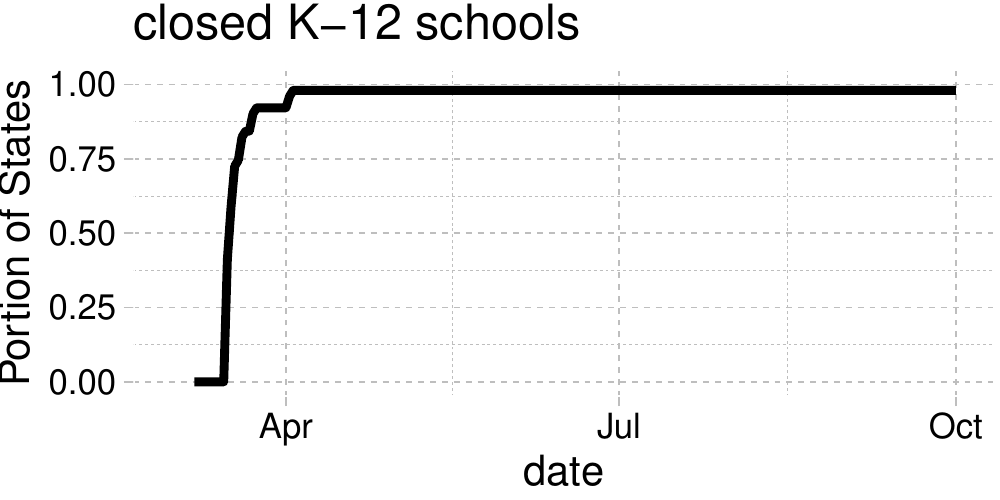}
      &
        \includegraphics[width=0.45\textwidth]{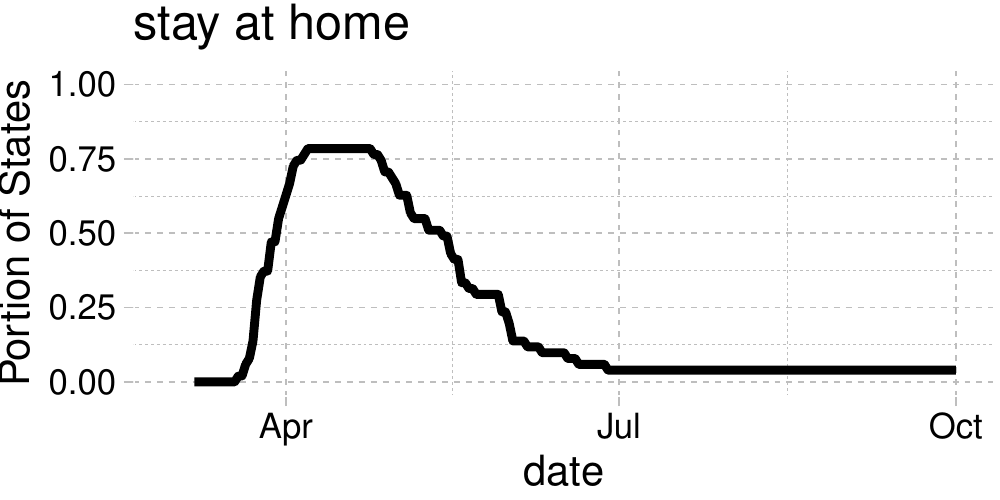}
      \\
      \includegraphics[width=0.45\textwidth]{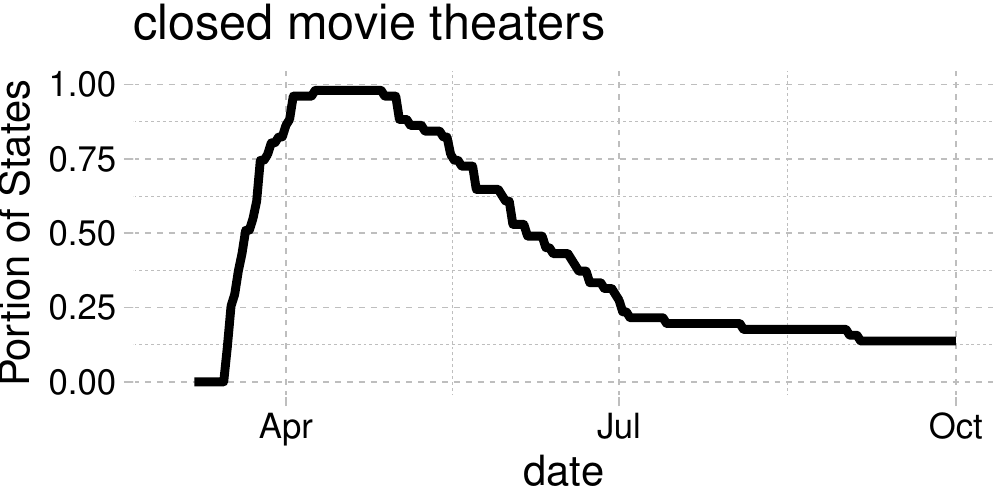}
      &
        \includegraphics[width=0.45\textwidth]{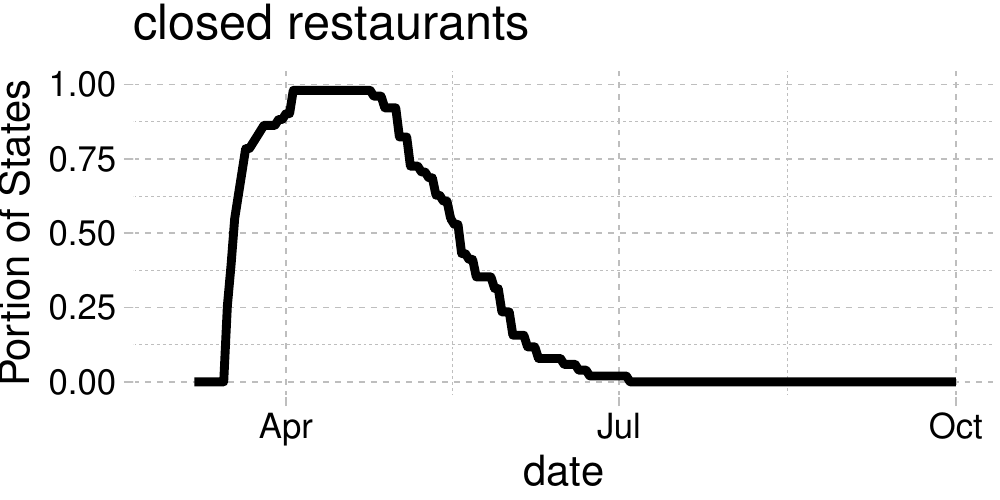}\\
        \includegraphics[width=0.45\textwidth]{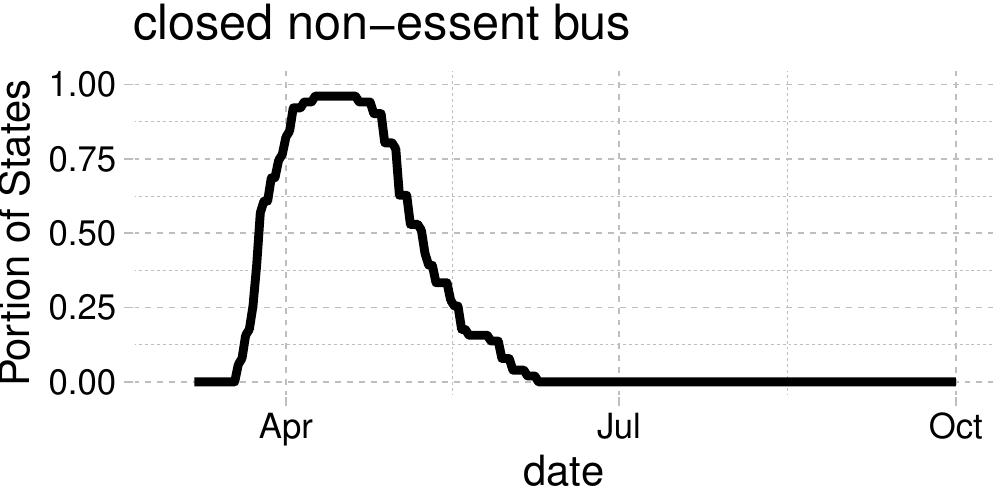}
    \end{tabular}
  \end{minipage}
\end{figure}

Figure \ref{fig:policyportion} presents the evolution of a portion of states with each policy for the updated data. When analyzing the impact of state-level policies, it is very important
to be aware of the range of policy variations observed in the data. 
For example, all states closed schools nearly the same time, so it is
impossible to reliably estimate the effect of school closures from
variation across states. In the case of public versus business mask
mandates, it is essential to realize that (except in Delaware for 3
days), states with a public mask mandate always also had a business
mask mandate. The set of observed mask policies across states are (i)
no mask mandate (ii) mask mandate only for businesses, and (iii) both
business and public mask mandate.

You can estimate a regression with separate indicators for business
and public mask mandates, but the interpretation of the coefficients
will not be straightforward. The coefficient on public mask mandate is
not the effect of a public mask mandate by itself. Estimating such an
effect is impossible because we never observe only a public mask
mandate. Instead, the coefficient on public mask mandate is the added
effect of a public mask mandate on top of a business mask mandate. If
you want to compare states with public mask mandates (which always
also have a business mask mandate) to states with no mask mandates,
then you should look at the sum of the coefficients on business mask
mandates and public mask mandates. Alternatively, you can estimate the
model with the mask policies encoded to represent the mask mandate
combinations that we observe. We construct a variable ``mask for employee only'' by assigning the value of one to state-day observation for which mask for the employee is implemented but masks in public spaces are not mandated. ``mask in public space'' variable takes one if mask mandate in public space is in place (which implies mask mandate for the employee for almost all cases except for 3 days lag for Vermont). We did regression analysis by including the smoothed version of these two variables in place of ``mask for employee'' variable in the original paper.

Table \ref{tab:PtoY-1} presents the regression result we obtained using the original data.  Both coefficients of ``mask for employee only'' and ``mask in public space''  variables are significant and substantial.


\begin{table}[!htbp] \centering
 \caption{\label{tab:PtoY-1}
 Original Data (Sample Period: March 7-June 3, 2020)}\vspace{-0.3cm}
\hspace{-3cm}   \resizebox{0.5\textwidth}{!}{
   \centering
   \tiny 
   \begin{minipage}{0.49\linewidth}
     \centering
     \input{tables_and_figures/cases-maskall-piy}
   \end{minipage} 
   }
\end{table}

\pagebreak

\subsection{The sensitivity of estimates with respect to the updated data}
Another important criticism of Lemoine is that our results change
substantially when using revised data.  

As shown in Table \ref{tab:PtoY-2}, when we use the updated data with the same sample period from March 7 to June 3, 2021, the
coefficients on masks are often smaller. This is consistent with what Lemoine reported in his post.

 \begin{table}[!htbp] \centering
 \caption{\label{tab:PtoY-2}
 Revised Data \textbf{with Hawaii's mask policy miscoded}  (Sample Period: March 7-June 3, 2020)}\vspace{-0.3cm}
\hspace{-3cm}   \resizebox{0.5\textwidth}{!}{
   \centering
   \tiny 
   \begin{minipage}{0.49\linewidth}
     \centering
     \input{tables_and_figures/cases-maskall-revised-piy}
   \end{minipage} 
   }
\end{table}

We wondered what caused the results to change. As a diagnostic, we computed the influence of each observation on the coefficient of mask policies, and we find that Hawaii is a notable high influence state along with Vermont, Connecticut, Montana, North Dakota, and West Virginia. We re-examine the updated data on mask policies for these 6 states by checking the state government websites. Then, we find the updated data substantially miscoded the date on which mask mandates began in Hawaii.  We also find business mask mandates for North Dakota are miscoded. See the appendix for details. 

Table \ref{tab:PtoY-3} shows the estimation result when we use the updated data after correcting the dates for mask mandates in Hawaii and North Dakota while keeping the same sample period. Once again, the effect of mask mandates is substantial and
statistically significant. Unlike the original data,  a public
mask mandate does not have a larger point estimate than a business mask mandate. However, in either case, the confidence intervals overlap
substantially. As we show below, the estimated coefficient of a public mask mandate becomes much larger than that of a business mask mandate when we use the sample of the extended period.

 \begin{table}[!htbp] \centering
 \caption{\label{tab:PtoY-3}
 Revised Data \textbf{with Hawaii's mask policy correctly coded}  (Sample Period: March 7-June 3, 2020)}\vspace{-0.3cm}
\hspace{-3cm}   \resizebox{0.5\textwidth}{!}{
   \centering
   \tiny 
   \begin{minipage}{0.49\linewidth}
     \centering
     \input{tables_and_figures/cases-maskall-revised-correct-piy}
   \end{minipage} 
   }
\end{table}

 \begin{table}[!htbp] \centering
 \caption{\label{tab:PtoY-4}
 Revised Data \textbf{without Hawaii } (Sample Period: March 7-June 3, 2020)}\vspace{-0.3cm}
\hspace{-3cm}   \resizebox{0.5\textwidth}{!}{
   \centering
   \tiny 
   \begin{minipage}{0.49\linewidth}
     \centering
     \input{tables_and_figures/cases-maskall-revised-nohawaii-piy}
   \end{minipage} 
   }
\end{table}

 \begin{table}[!htbp] \centering
 \caption{\label{tab:PtoY-6}
 Extended  Revised Data     (Sample Period: March 7-\textbf{August 1}, 2020)}\vspace{-0.3cm}
\hspace{-3cm}   \resizebox{0.5\textwidth}{!}{
   \centering
   \tiny 
   \begin{minipage}{0.49\linewidth}
     \centering
     \input{tables_and_figures/cases-maskall-long-aug-piy}
   \end{minipage} 
   }
\end{table}

Given the concerns about the significant influence of Hawaii, one could
consider dropping it from the sample. Table  \ref{tab:PtoY-4} presents the estimates when we use the updated data after dropping Hawaii from the sample. The results are mainly similar to those in Table \ref{tab:PtoY-3}.

\subsection{Extending the sample period in the updated data}

As shown in Figure \ref{fig:policyportion}, a relatively small portion of states has implemented mask mandates in public spaces in the middle of May. This made it difficult to identify the coefficient of  ``masks in public spaces'' using the sample from March 7 to June 3 of 2020. 

We fix the end of the sample period  to  June 3, 2020 when we worked on revising our paper for publication because constantly updating the data would make the revision process very difficult.  A preprint version of our paper was available in ArXiv in late May of 2020 and was submitted
for publication shortly after, which is why we did not analyze either the updated data or the extended data in our original paper. 

Now we have access to a more extended sample period, and we can examine how extending the sample period changes the estimates. Analyzing the data with a longer sample period has the advantage of better identifying the coefficient of mask mandates in public spaces variable because many states implemented public mask mandates in June and July.

Here, we consider three different end dates for the sample: July 1, August 1, and September 1 of 2020. Note that the data set does not contain the data on school re-opening dates in the fall of 2020---it is not easy to obtain a single school opening date at the state level  because the school opening dates vary across school districts within each state; furthermore, their modes of school openings (e.g., in-person vs. online) differ across school districts \citep{Chernozhukov2021pnas}. In addition, there is almost no change in the recorded policies in the data set after September 1 in Figure \ref{fig:policyportion}.
For this reason, we believe that the data set up to August 1 before none of the school districts has opened their schools may be a good data set to examine. 

Table  \ref{tab:PtoY-6} presents the estimation results using the sample ending on August 1, while Tables \ref{tab:PtoY-5}-\ref{tab:PtoY-7} in the appendix show the estimates using the sample ending on July 1 and September 1, respectively.

Across all specifications and samples in Tables  \ref{tab:PtoY-6}-\ref{tab:PtoY-7}, the estimated coefficients of  ``masks in public spaces'' range from $-0.097$ to $-0.124$, and   are roughly twice as large as those of ``masks for employees only."  Therefore, once the sample period is extended, the estimates indicate that the effect of the public mask mandates is larger than that of the business mask mandates. Confidence intervals get narrower as we expand the sample period, suggesting that additional variation in the  mask mandate policies in June and July helps identifying the coefficients of mask policies better.r.

\subsection{Estimation using Simulated Data}

\begin{figure}[!htbp]\caption{Estimated Effects of Non-Pharmaceutical Intervention on the Simulated Data from Lemoines' Blog  \label{fig:simulation-1}}\medskip 
  \begin{minipage}{0.9\linewidth} 
      \includegraphics[width=\textwidth]{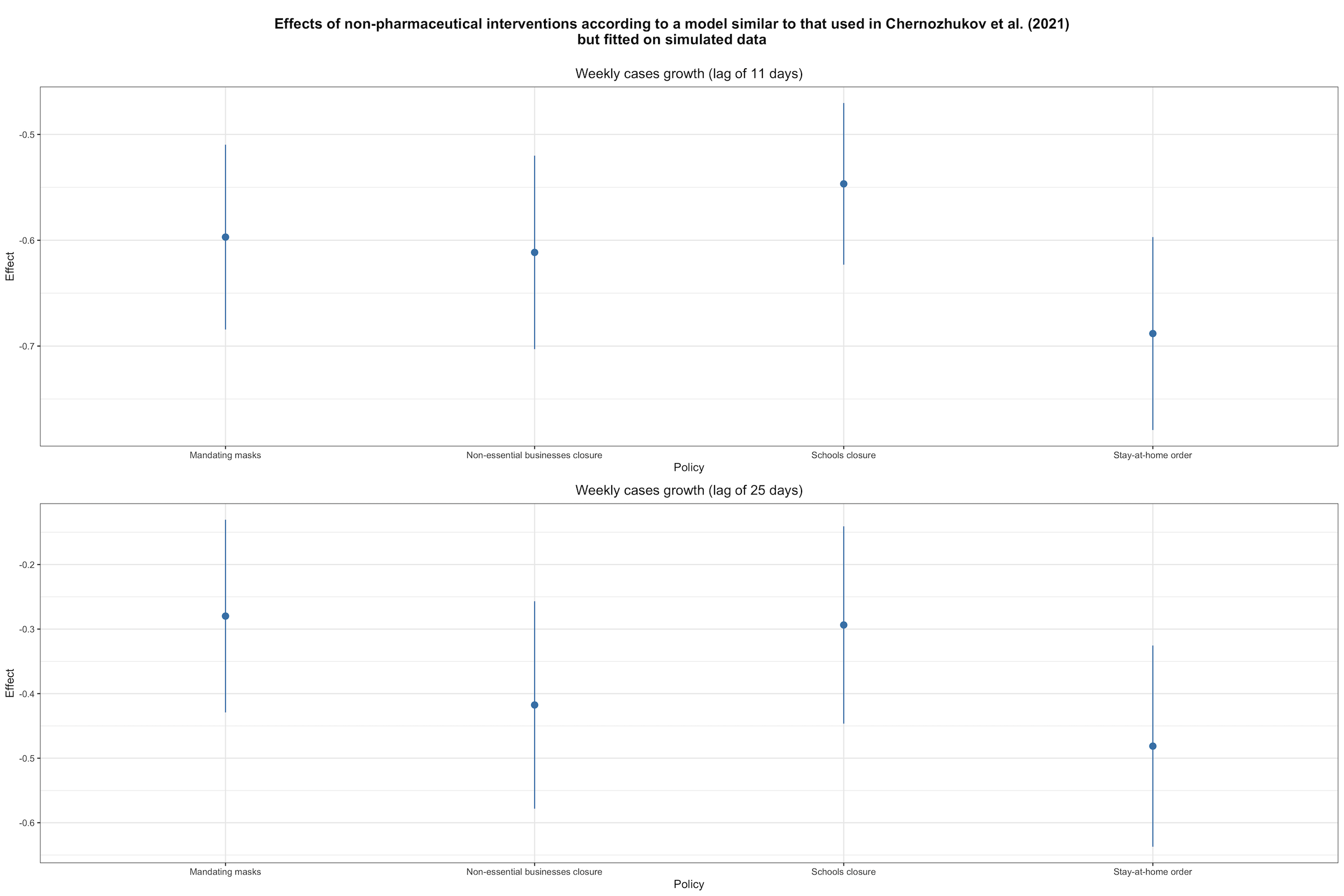} 
  \end{minipage}
\end{figure}  

Lemoine simulated data from the SIR model, where a basic reproduction number is set to 2.5 and each of randomly drawn 4 policies (“mandating masks”, “schools closure”, “stay-at-home order” and “non-essential businesses closure”) additively reduces the effective reproduction number by slightly more than 0.5. He generated 1,500 epidemics and randomly selected 50 of those epidemics among those for which the final attack rate was between 0.5\% and 10\%. He estimated our regression models using these selected simulated data. 

Lemoines plots the estimated coefficients of four policies using the weekly case growth rate with the lag of 11 days or 25 days as shown in Figure \ref{fig:simulation-1}. All of the estimated coefficients are negatively estimated. Thus, these regression results on simulated data suggest that our regression model correctly captures the impact of non-pharmaceutical interventions. In our view, this indicates that the regression-based procedure can correctly detect the effect of non-pharmaceutical interventions from the data. 

\begin{figure}[!htbp]\caption{Effects of Removing Policies from Lemoines' Blog  \label{fig:simulation-2}}\medskip 
  \begin{minipage}{1.3\linewidth} 
      \includegraphics[width=\textwidth, trim=2cm 15cm 0 2.5cm]{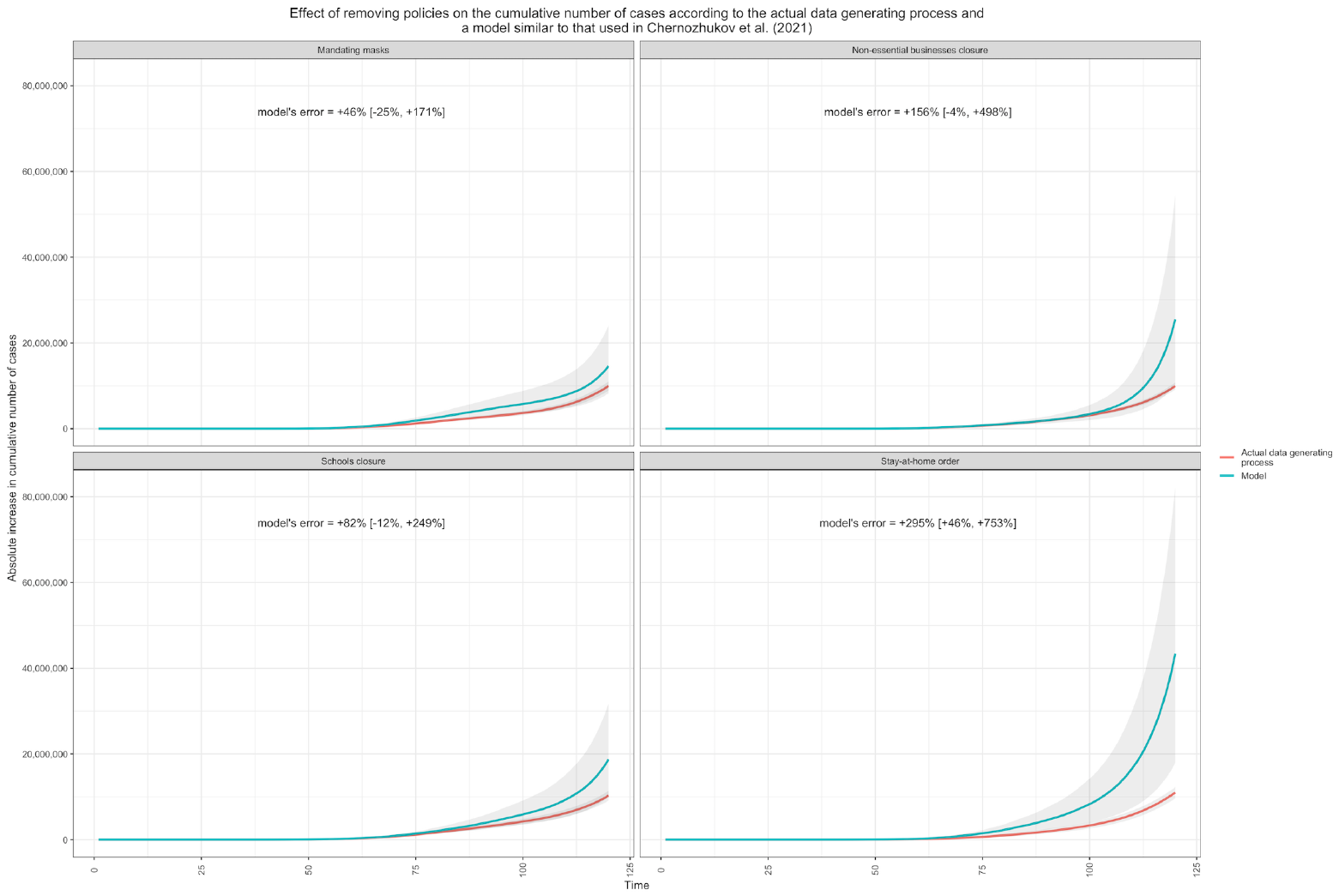} 
  \end{minipage}
\end{figure}

Lemoines also computed the counterfactual effect given the estimate, and compare it with the ``confidence interval'' he obtained from repeatedly simulating his SIR model and computing the counterfactual effect implied by the SIR model. His result on case growth regression is presented in Figure \ref{fig:simulation-2}. In Figure \ref{fig:simulation-2}, the  counterfactual effects estimated by the regression model is within the SIR model-based  ``confidence interval'' except for stay-at-home order. Lemoines characterizes this observation as an indication that ``the kind of model used by Chernozhukov et al. performs extremely poorly.''   We are not sure if his assertion that our model performs ``extremely poorly'' is accurate given that the estimates are within his model-based ``confidence intervals.''

Lemoines initially presented a version of this figure that extends the simulated period to 200+ days rather than 120 days, which gives an impression that the  estimated regression-based model  performs poorly because the longer the time-horizon you consider, the model error will accumulate more in the presence of exponential growth, giving poorer fit. Later, Lemoines replaced this figure with the current version without indicating that he updated this figure in his blog. His statement that our model ``performs extremely poorly'' might reflect on his previously presented figure, which he has replaced. The length of the actual data we use in our paper is 96 days and so considering a shorter period than 120 days would be more appropriate. Contrary to his assertion, if Figure \ref{fig:simulation-2} were ended at 96 days, it may give an impression that the counterfactual effects estimated by our model may not be so different from the ones produced by the simulated model.

His SIR model does not incorporate voluntary behavioral changes toward the transmission risk, which is an essential part of our regression model. In our model specification, the case growth rate depends on the past level and growth of cases to capture the fact that people voluntarily change their behavior in response to the risk of infections. This seems an important omission in his SIR model given the context of his critique. In his model, he sets the value of basic reproduction number to 2.5 and, without voluntary behavioral change, his simulated model tends to generate the high proportion of people who have ever been infected within a short period.

\subsection{Placebo test}

Lemoine claims that he did a placebo test and easily found spurious effects. His placebo test is based on the simulated data based on his SIR model.\footnote{His blog explains his placebo test as follows:
\begin{quote}
 I fitted the model on the data that I simulated above, but added a policy that did not actually play any role in the data generating process whose starting time and duration were drawn from a random distribution. I repeated this procedure 500 times for both weekly cases growth and weekly deaths growth. 
\end{quote}}

We also implemented our own placebo test to examine if we can find spurious effects. Specifically, we generated 500 samples, where each sample is identical to the original sample except that state-level mask variables (i.e., ``masks in public spaces'' and ``masks for employees only") are replaced with those of other states that are randomly selected through permutation. Because each state's mask variables are replaced with randomly selected other state's mask variables, we expect that the estimated coefficients of mask variables would be centered around zero if there is no spurious effects arising from, for example, aggregate time effect.

\begin{figure}[!htbp]\caption{Estimated Coefficients under Randomly Permutated Mask Policies \label{fig:placebo}}\medskip 
  \begin{minipage}{\linewidth} 
\begin{tabular}{cc}
\small\textbf{(a) Sample Period: March 7-June 3, 2020}&\small\textbf{(b) Sample Period: March 7-August 1, 2020}\\
      \includegraphics[width=0.5\textwidth]{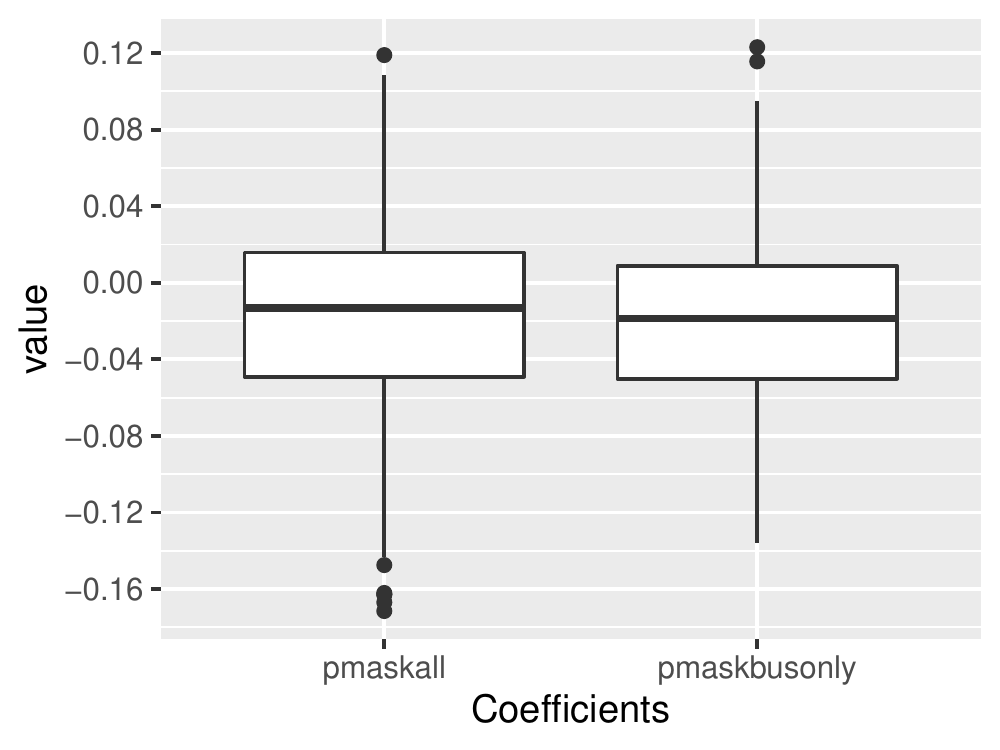} &      \includegraphics[width=0.5\textwidth]{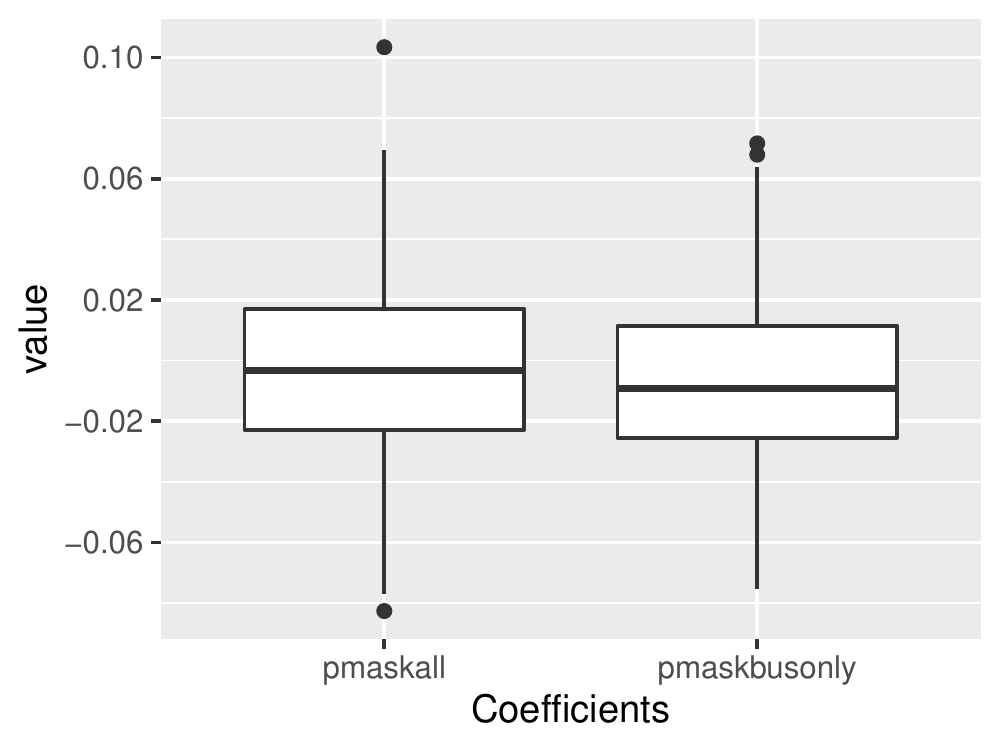}
\end{tabular} \centering
  \end{minipage}
\end{figure}

Figure \ref{fig:placebo}(a) presents the result of our placebo test when we use the sample from March 7 to June 3, 2020. The estimates are roughly centered around -0.02 for both mask variables while the inter-quartile ranges for  ``masks in public spaces'' and ``masks for employees only" are around $[-0.05,0.02]$. In Figure \ref{fig:placebo}(b), when we use the extended sample from March 7 to August 1, 2020, the estimates are  centered around $-0.01$ while the inter-quartile ranges are narrower with  $[-0.03,0.02]$. Overall, the result of our placebo test indicates that there won't be any systematic spurious effects we may find from our regression analysis, especially if we use the data set with the extended sample period to August 1, 2020.

\subsection{Other critiques}

Lemoine also asserts that our MA(7) “smoothing” is “inappropriate” without any justification.  MA(7) smoothing of outcomes (deaths and case counts) is made routinely in statistical analysis of data at daily frequency – this is done to remove very sharp “day” seasonality and high noise for death counts. For example IHME’s, NYT, FT standard display of cases and death cases involves showing MA(7)-smoothed series.  On the other hand, MA(7)-smoothing of policies or other explanatory variables  X is a simple approach to modeling outcome as a response to X when the time-to-response is stochastic.  For example, a recorded death can be a consequence of an infection that happened to 20, 21, 22,.., days ago.  When it is convenient to support his arguments, Lemoine  acknowledges this stochastic time-to-response in his footnotes but oddly does not use this point to support our argument.
Furthermore, when the outcome is MA(7) smoothed, using smoothed policies makes even more sense, even when the response-time is deterministic. In this case, by construction, the outcome is a response to a moving average of policies. In summary, using unsmoothed data can only makes sense if we artificially try to lower the precision of the statistical inference, and this seems what Lemoine tries to achieve (given his ideological stance). Therefore, this point simply does not make sense.  

Lemoine's critique also concerns the width of the confidence intervals. For example, in a counterfactual simulation of the dynamic effect of business closures in our basic model, the 90\% confidence interval for confirmed death counts is from [-4\%, +64\%] (the confidence intervals for case counts are somewhat tighter). Lemoine concludes, therefore, that business closures had zero effect because the null hypothesis of zero effect cannot be rejected at 10\% significance.  This is an incorrect interpretation of confidence intervals.  Confidence intervals collect all point hypotheses that can not be rejected; in particular, we can not rule out large or no effects on deaths of the mitigation policies (for case counts, the 90\% Cis based on baseline model was ruling out zero effects).  In summary, this point seems to be another honest mistake.  

Furthermore, when we use the extended sample period to, say, August 1 of 2020, the coefficients of mask policy variables are more precisely estimated with the shorter confidence interval. If he had a concern over the wide confidence interval, Lemoine should have analyzed the updated data with the extended sample period when he investigated the updated data rather than keeping the original sample period of March 7--June 3 of 2020.  We may obtain the shorter confidence interval by using the more extended sample period in this context because many states implemented mask mandates in June and July, helping identify the mask coefficients.

%


There are other points he raised, but we simply did not have time to review all the points and focused here on what we think as the main ones.

\section{Appendix}

\subsection{Influence}

 Lemoine show that our results change
substantially when using revised data.
What caused the results to change? One diagnostic is to look at the
influence of each observation on the coefficients on the mask
policies. Here we plot $$ \frac{-1}{1 - x_{it}' ({X}'{X})^{-1} x_{it}}
({X}'{X})^{-1} {pmask}_{it} \hat{\epsilon}_{it} $$ which is how much
the coefficient would change if that observation were deleted. A state
with positive influence values means that state contributes to the
estimate being negative, and vice versa.

Figure 2 plots the influence of each state-day observation on the  estimated coefficient on mask for business only variables for the original data and the updated data for the sample period of March 7-June 3, 2021, while Figure 3 plots the corresponding figures for the public mask coefficients. Influential state observations include those of Hawaii, Vermont, Connecticut, Montana, North Dakota, and West Virginia. We re-examine the updated data on mask policies for these 6 states by checking the state government websites. 

 The old data has Hawaii
beginning both business and public mask mandates on April
16, 2020. The new data lists business mask mandates beginning on June
11, 2020 and public mandates on November 16, 2020. In reality, the
Hawaii [governor proclamation on April 16,
2020](https://governor.hawaii.gov/wp-content/uploads/2020/04/2004088-ATG\_Fifth-Supplementary-Proclamation-for-COVID-19-distribution-signed.pdf)
*recommended* face masks in public and mandated face masks for
employess and customers in essential businesses. The same requirements
are repeated in [a proclamation on June 10,
2020.](https://governor.hawaii.gov/wp-content/uploads/2020/06/2006097A-ATG\_Ninth-Supplementary-Proclamation-COVID-19-distribution-signed.pdf). A
[proclamation on November
16](https://governor.hawaii.gov/wp-content/uploads/2020/11/2011051-ATG\_Fifteenth-Proclamation-Related-to-the-COVID-19-Emergency-distribution-signed.pdf)
extends the mask mandate to all persons in public. Thus, it appears
neither the old or new data is entirely correct. The original coding
of the data is understandable --- the order on April
16, 2020 is stronger than requiring masks for employees in businesses,
since it includes customers too. The updated data appears to just be
a mistake. 

We also check mask policies for the following states that have high influence on mask coefficients: Vermont, Connecticut, Montana, North Dakota, and West Virginia. Then, we find that  business mask mandates for North Dakota were announced on April 29 and then became effective on May 1 while the revised data has business mask mandate in North Dakota beginning on April 28 (https://www.governor.nd.gov/executive-orders and https://www.governor.nd.gov/sites/www/files/documents/executive-orders/Executive\%20Order\%202020-06.4.pdf).   In the revised data,  West Virginia starts its universal public mask mandates on July 7 of 2020 but this public mask mandates only covers when social distancing is not possible. We keep West Virginia's starting date for public mask mandates as it is.

\begin{figure}[!htbp]\caption{Influence of each state-day observation on  pmaskbusonly coefficient for the original and the updated data \label{fig:influence1}}\medskip   
  \begin{minipage}{0.7\linewidth}
    \begin{tabular}{c}
      \includegraphics[width=\textwidth]{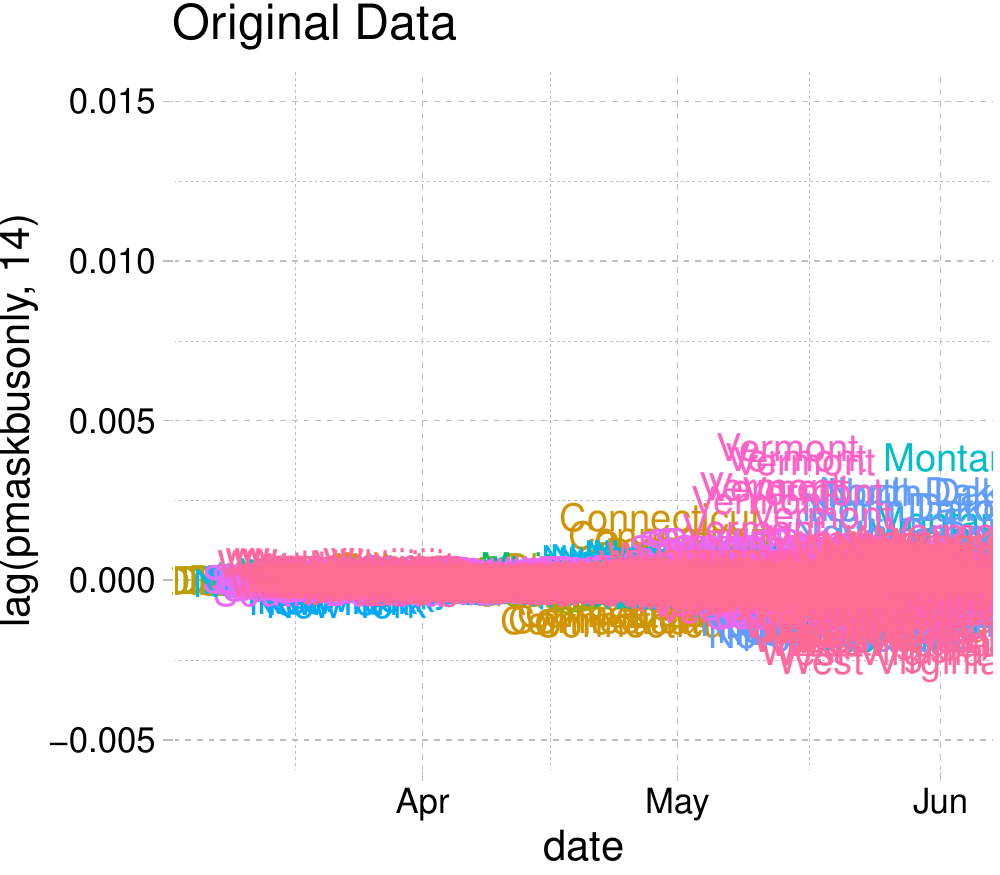}\\
        \includegraphics[width=\textwidth]{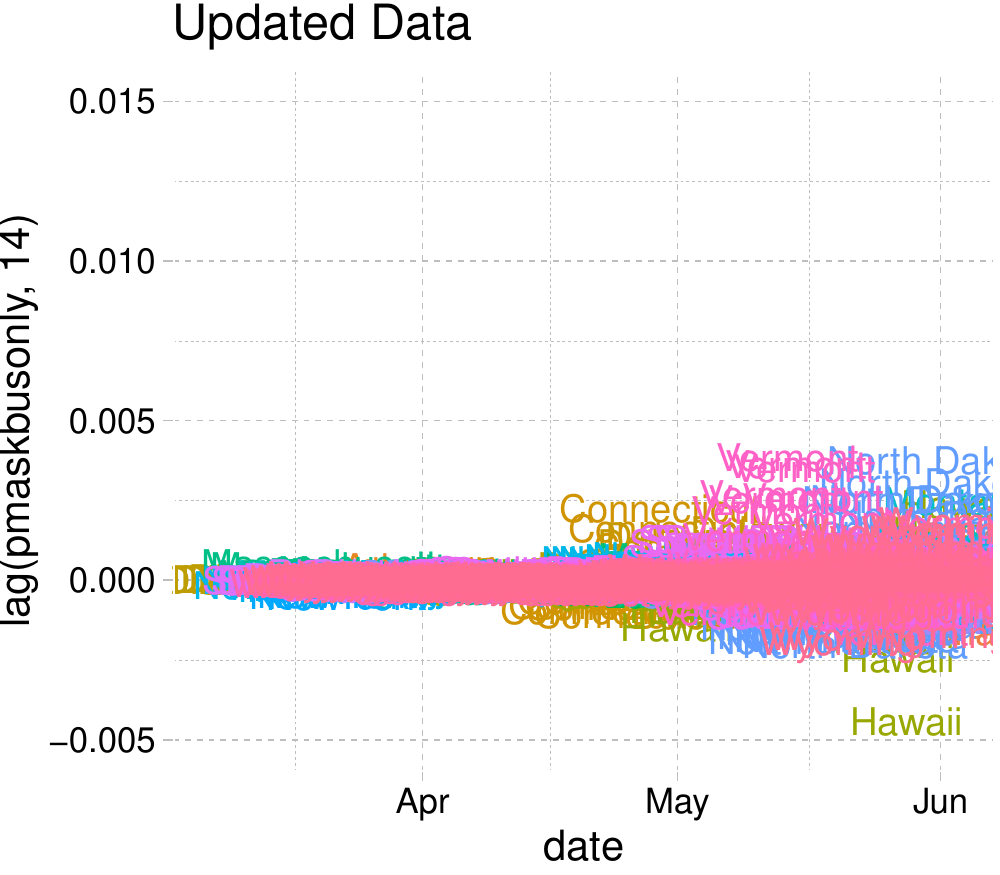}
    \end{tabular} 
    \end{minipage}
\end{figure}

\begin{figure}[!htbp]\caption{Influence of each state-day observation on  pmaskall coefficient for the original and the updated data \label{fig:influence}}\medskip 
  \begin{minipage}{0.7\linewidth}
    \begin{tabular}{c}
      \includegraphics[width=\textwidth]{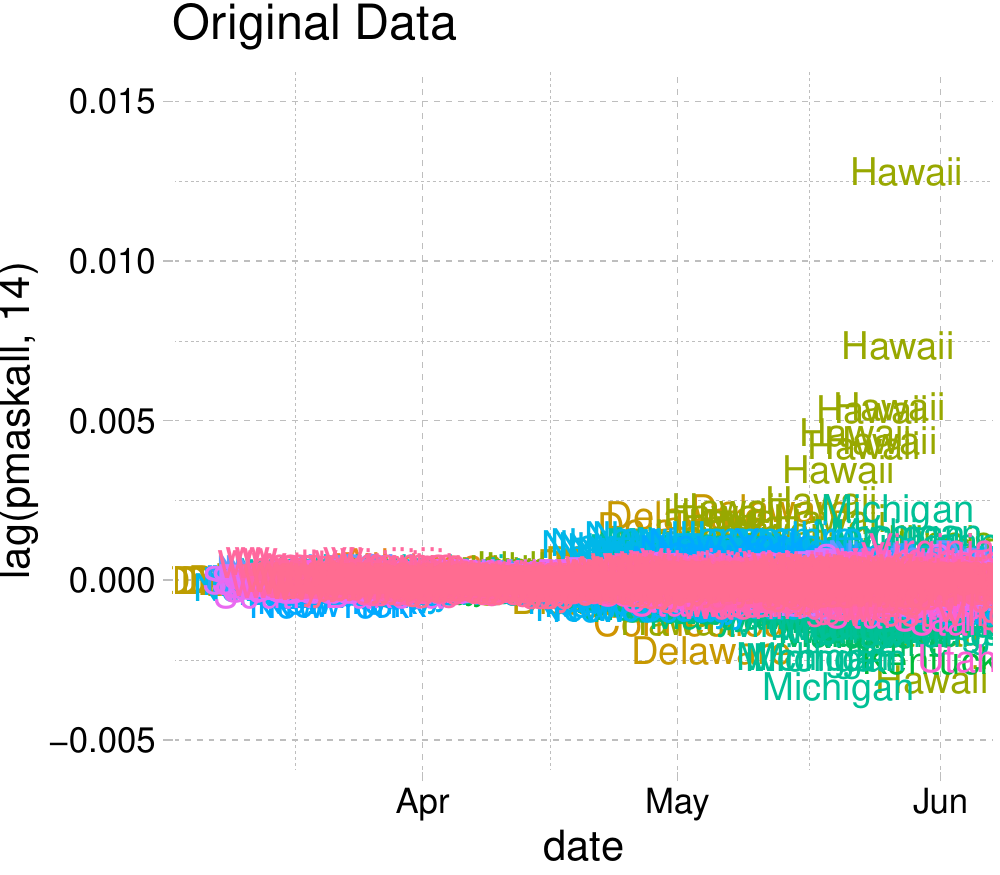}\\ 
        \includegraphics[width=\textwidth]{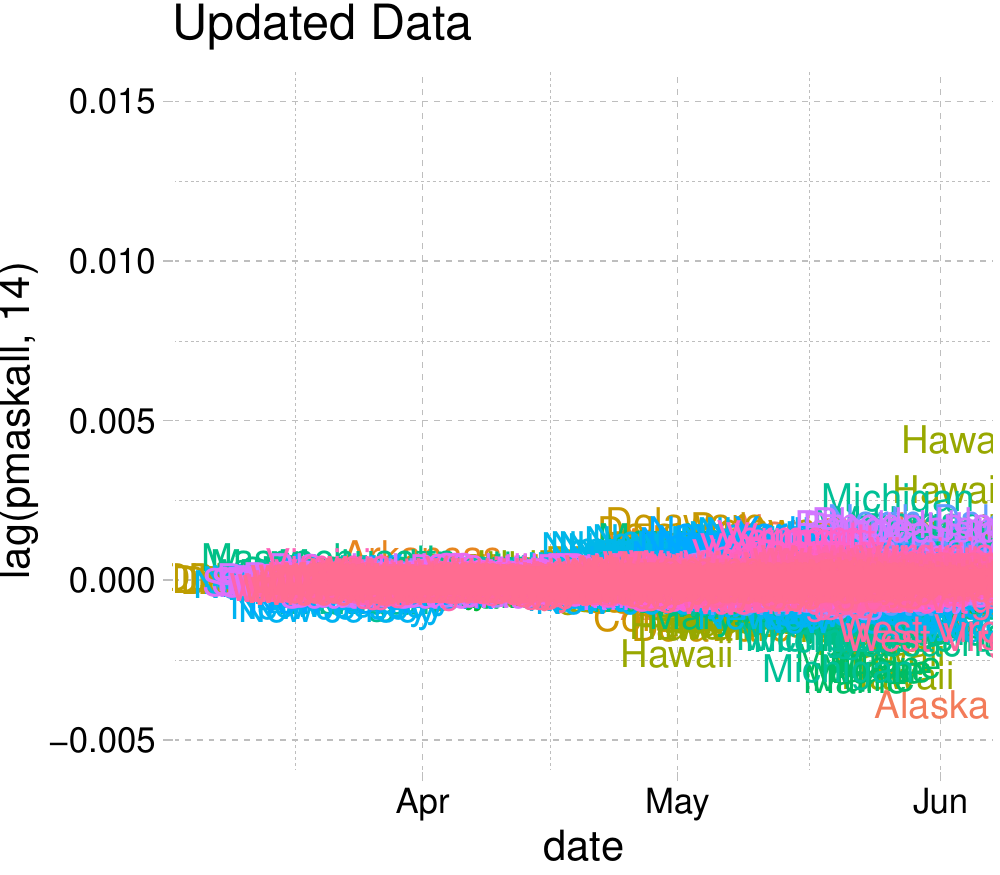} 
    \end{tabular} 
    \end{minipage}
\end{figure}

\subsection{Additional Tables}

\begin{table}[!htbp] \centering
 \caption{\label{tab:PtoY-5}
 Extended Revised Data    (Sample Period: March 7-\textbf{July 1}, 2020)}\vspace{-0.3cm}
\hspace{-3cm}   \resizebox{0.5\textwidth}{!}{
   \centering
   \tiny 
   \begin{minipage}{0.49\linewidth}
     \centering
     \input{tables_and_figures/cases-maskall-long-july-piy}
   \end{minipage} 
   }
\end{table}

 \begin{table}[!htbp] \centering
 \caption{\label{tab:PtoY-7}
 Extended Revised Data   (Sample Period: March 7-\textbf{September 1}, 2020)}\vspace{-0.3cm}
\hspace{-3cm}   \resizebox{0.5\textwidth}{!}{
   \centering
   \tiny 
   \begin{minipage}{0.49\linewidth}
     \centering
     \input{tables_and_figures/cases-maskall-long-sept-piy}
   \end{minipage} 
   }
\end{table}

\bibliographystyle{jpe}
\bibliography{covid}

\end{document}

%% file: tables_and_figures/cases-maskall-piy.tex
\begin{tabular}{@{\extracolsep{1pt}}lcccc} 
\\[-1.8ex]\hline 
\hline \\[-1.8ex] 
 & \multicolumn{4}{c}{\textit{Dependent variable:}} \\ 
\cline{2-5} 
 & \multicolumn{4}{c}{$\Delta \log \Delta C_{it}$} \\ 
\\[-1.8ex] & (1) & (2) & (3) & (4)\\ 
\hline \\[-1.8ex] 
 lag(masks for employees only, 14) & $-$0.061$^{*}$ & $-$0.062$^{*}$ & $-$0.082$^{***}$ & $-$0.085$^{***}$ \\ 
  & (0.037) & (0.037) & (0.031) & (0.032) \\ 
  lag(masks in public spaces, 14) & $-$0.147$^{*}$ & $-$0.140$^{*}$ & $-$0.171$^{**}$ & $-$0.163$^{**}$ \\ 
  & (0.076) & (0.080) & (0.067) & (0.069) \\ 
  lag(closed K-12 schools, 14) & $-$0.231$^{**}$ & $-$0.235$^{**}$ & 0.043 & 0.036 \\ 
  & (0.095) & (0.102) & (0.107) & (0.112) \\ 
  lag(stay at home, 14) & $-$0.137$^{**}$ & $-$0.135$^{**}$ & $-$0.122$^{**}$ & $-$0.114$^{**}$ \\ 
  & (0.061) & (0.057) & (0.056) & (0.052) \\ 
  lag(business closure policies, 14) & $-$0.075 &  & 0.009 &  \\ 
  & (0.067) &  & (0.061) &  \\ 
  lag(closed movie theaters, 14) &  & 0.007 &  & 0.049 \\ 
  &  & (0.054) &  & (0.047) \\ 
  lag(closed restaurants, 14) &  & $-$0.041 &  & $-$0.010 \\ 
  &  & (0.049) &  & (0.044) \\ 
  lag(closed non-essent bus, 14) &  & $-$0.035 &  & $-$0.020 \\ 
  &  & (0.047) &  & (0.041) \\ 
  lag($\Delta \log \Delta C_{it}$, 14) & 0.042$^{*}$ & 0.043$^{*}$ & 0.036 & 0.036 \\ 
  & (0.024) & (0.024) & (0.028) & (0.028) \\ 
  lag($\log \Delta C_{it}$, 14) & $-$0.135$^{***}$ & $-$0.136$^{***}$ & $-$0.084$^{***}$ & $-$0.084$^{***}$ \\ 
  & (0.025) & (0.026) & (0.031) & (0.031) \\ 
  lag($\Delta \log \Delta C_{it}$.national, 14) &  &  & $-$0.128$^{***}$ & $-$0.124$^{***}$ \\ 
  &  &  & (0.039) & (0.042) \\ 
  lag($\log \Delta C_{it}$.national, 14) &  &  & $-$0.259$^{***}$ & $-$0.260$^{***}$ \\ 
  &  &  & (0.049) & (0.049) \\ 
  $\Delta \log T_{it}$ & 0.111$^{***}$ & 0.111$^{***}$ & 0.115$^{***}$ & 0.115$^{***}$ \\ 
  & (0.028) & (0.028) & (0.026) & (0.026) \\ 
 \hline \\[-1.8ex] 
state variables & Yes & Yes & Yes & Yes \\ 
Month $\times$ state variables & Yes & Yes & Yes & Yes \\ 
\hline \\[-1.8ex] 
$\sum_j \mathrm{Policy}_j$ & -0.651$^{***}$ & -0.641$^{***}$ & -0.324$^{*}$ & -0.307 \\ 
 & (0.203) & (0.211) & (0.183) & (0.188) \\ 
Observations & 3,807 & 3,807 & 3,807 & 3,807 \\ 
R$^{2}$ & 0.749 & 0.749 & 0.763 & 0.764 \\ 
Adjusted R$^{2}$ & 0.746 & 0.746 & 0.761 & 0.761 \\ 
\hline 
\hline \\[-1.8ex] 
\textit{Note:}  & \multicolumn{4}{r}{$^{*}$p$<$0.1; $^{**}$p$<$0.05; $^{***}$p$<$0.01} \\ 
\end{tabular} 

%% file: tables_and_figures/cases-maskall-revised-piy.tex
\begin{tabular}{@{\extracolsep{1pt}}lcccc} 
\\[-1.8ex]\hline 
\hline \\[-1.8ex] 
 & \multicolumn{4}{c}{\textit{Dependent variable:}} \\ 
\cline{2-5} 
 & \multicolumn{4}{c}{$\Delta \log \Delta C_{it}$} \\ 
\\[-1.8ex] & (1) & (2) & (3) & (4)\\ 
\hline \\[-1.8ex] 
 lag(masks for employees only, 14) & $-$0.062 & $-$0.061 & $-$0.082$^{***}$ & $-$0.085$^{***}$ \\ 
  & (0.040) & (0.040) & (0.031) & (0.030) \\ 
  lag(masks in public spaces, 14) & $-$0.039 & $-$0.036 & $-$0.085$^{**}$ & $-$0.087$^{**}$ \\ 
  & (0.050) & (0.050) & (0.043) & (0.043) \\ 
  lag(closed K-12 schools, 14) & $-$0.205$^{***}$ & $-$0.205$^{***}$ & $-$0.068 & $-$0.069 \\ 
  & (0.067) & (0.067) & (0.071) & (0.070) \\ 
  lag(stay at home, 14) & $-$0.097$^{*}$ & $-$0.098$^{*}$ & $-$0.093$^{*}$ & $-$0.091$^{*}$ \\ 
  & (0.051) & (0.052) & (0.049) & (0.050) \\ 
  lag(business closure policies, 14) & $-$0.039 &  & 0.030 &  \\ 
  & (0.055) &  & (0.055) &  \\ 
  lag(closed movie theaters, 14) &  & 0.001 &  & 0.022 \\ 
  &  & (0.059) &  & (0.062) \\ 
  lag(closed restaurants, 14) &  & $-$0.038 &  & 0.012 \\ 
  &  & (0.054) &  & (0.057) \\ 
  lag(closed non-essent bus, 14) &  & $-$0.001 &  & $-$0.003 \\ 
  &  & (0.037) &  & (0.031) \\ 
  lag($\Delta \log \Delta C_{it}$, 14) & 0.067$^{***}$ & 0.068$^{***}$ & 0.055$^{**}$ & 0.054$^{**}$ \\ 
  & (0.024) & (0.025) & (0.027) & (0.027) \\ 
  lag($\log \Delta C_{it}$, 14) & $-$0.134$^{***}$ & $-$0.135$^{***}$ & $-$0.088$^{***}$ & $-$0.088$^{***}$ \\ 
  & (0.020) & (0.021) & (0.027) & (0.027) \\ 
  lag($\Delta \log \Delta C_{it}$.national, 14) &  &  & $-$0.108$^{***}$ & $-$0.109$^{**}$ \\ 
  &  &  & (0.041) & (0.046) \\ 
  lag($\log \Delta C_{it}$.national, 14) &  &  & $-$0.215$^{***}$ & $-$0.215$^{***}$ \\ 
  &  &  & (0.040) & (0.040) \\ 
  $\Delta \log T_{it}$ & 0.171$^{***}$ & 0.170$^{***}$ & 0.166$^{***}$ & 0.166$^{***}$ \\ 
  & (0.047) & (0.047) & (0.043) & (0.043) \\ 
 \hline \\[-1.8ex] 
state variables & Yes & Yes & Yes & Yes \\ 
Month $\times$ state variables & Yes & Yes & Yes & Yes \\ 
\hline \\[-1.8ex] 
$\sum_j \mathrm{Policy}_j$ & -0.442$^{***}$ & -0.437$^{***}$ & -0.299$^{**}$ & -0.302$^{**}$ \\ 
 & (0.119) & (0.121) & (0.134) & (0.137) \\ 
Observations & 3,726 & 3,726 & 3,726 & 3,726 \\ 
R$^{2}$ & 0.740 & 0.740 & 0.752 & 0.752 \\ 
Adjusted R$^{2}$ & 0.738 & 0.738 & 0.749 & 0.749 \\ 
\hline 
\hline \\[-1.8ex] 
\textit{Note:}  & \multicolumn{4}{r}{$^{*}$p$<$0.1; $^{**}$p$<$0.05; $^{***}$p$<$0.01} \\ 
\end{tabular} 

%% file: tables_and_figures/cases-maskall-revised-correct-piy.tex
\begin{tabular}{@{\extracolsep{1pt}}lcccc} 
\\[-1.8ex]\hline 
\hline \\[-1.8ex] 
 & \multicolumn{4}{c}{\textit{Dependent variable:}} \\ 
\cline{2-5} 
 & \multicolumn{4}{c}{$\Delta \log \Delta C_{it}$} \\ 
\\[-1.8ex] & (1) & (2) & (3) & (4)\\ 
\hline \\[-1.8ex] 
 lag(masks for employees only, 14) & $-$0.141$^{***}$ & $-$0.143$^{***}$ & $-$0.144$^{***}$ & $-$0.150$^{***}$ \\ 
  & (0.042) & (0.042) & (0.031) & (0.030) \\ 
  lag(masks in public spaces, 14) & $-$0.072$^{*}$ & $-$0.071$^{*}$ & $-$0.112$^{***}$ & $-$0.116$^{***}$ \\ 
  & (0.042) & (0.042) & (0.041) & (0.041) \\ 
  lag(closed K-12 schools, 14) & $-$0.202$^{***}$ & $-$0.203$^{***}$ & $-$0.066 & $-$0.069 \\ 
  & (0.065) & (0.065) & (0.071) & (0.069) \\ 
  lag(stay at home, 14) & $-$0.095$^{*}$ & $-$0.094$^{*}$ & $-$0.093$^{**}$ & $-$0.088$^{*}$ \\ 
  & (0.049) & (0.050) & (0.047) & (0.049) \\ 
  lag(business closure policies, 14) & $-$0.053 &  & 0.017 &  \\ 
  & (0.052) &  & (0.051) &  \\ 
  lag(closed movie theaters, 14) &  & 0.009 &  & 0.025 \\ 
  &  & (0.060) &  & (0.063) \\ 
  lag(closed restaurants, 14) &  & $-$0.041 &  & 0.012 \\ 
  &  & (0.053) &  & (0.057) \\ 
  lag(closed non-essent bus, 14) &  & $-$0.019 &  & $-$0.019 \\ 
  &  & (0.039) &  & (0.031) \\ 
  lag($\Delta \log \Delta C_{it}$, 14) & 0.063$^{***}$ & 0.064$^{***}$ & 0.054$^{**}$ & 0.054$^{**}$ \\ 
  & (0.024) & (0.024) & (0.027) & (0.027) \\ 
  lag($\log \Delta C_{it}$, 14) & $-$0.135$^{***}$ & $-$0.136$^{***}$ & $-$0.092$^{***}$ & $-$0.091$^{***}$ \\ 
  & (0.019) & (0.020) & (0.024) & (0.025) \\ 
  lag($\Delta \log \Delta C_{it}$.national, 14) &  &  & $-$0.114$^{***}$ & $-$0.116$^{***}$ \\ 
  &  &  & (0.040) & (0.045) \\ 
  lag($\log \Delta C_{it}$.national, 14) &  &  & $-$0.212$^{***}$ & $-$0.214$^{***}$ \\ 
  &  &  & (0.039) & (0.039) \\ 
  $\Delta \log T_{it}$ & 0.168$^{***}$ & 0.168$^{***}$ & 0.163$^{***}$ & 0.164$^{***}$ \\ 
  & (0.046) & (0.046) & (0.043) & (0.043) \\ 
 \hline \\[-1.8ex] 
state variables & Yes & Yes & Yes & Yes \\ 
Month $\times$ state variables & Yes & Yes & Yes & Yes \\ 
\hline \\[-1.8ex] 
$\sum_j \mathrm{Policy}_j$ & -0.563$^{***}$ & -0.562$^{***}$ & -0.398$^{***}$ & -0.404$^{***}$ \\ 
 & (0.135) & (0.138) & (0.132) & (0.135) \\ 
Observations & 3,726 & 3,726 & 3,726 & 3,726 \\ 
R$^{2}$ & 0.742 & 0.742 & 0.754 & 0.754 \\ 
Adjusted R$^{2}$ & 0.740 & 0.740 & 0.751 & 0.751 \\ 
\hline 
\hline \\[-1.8ex] 
\textit{Note:}  & \multicolumn{4}{r}{$^{*}$p$<$0.1; $^{**}$p$<$0.05; $^{***}$p$<$0.01} \\ 
\end{tabular} 

%% file: tables_and_figures/cases-maskall-revised-nohawaii-piy.tex
\begin{tabular}{@{\extracolsep{1pt}}lcccc} 
\\[-1.8ex]\hline 
\hline \\[-1.8ex] 
 & \multicolumn{4}{c}{\textit{Dependent variable:}} \\ 
\cline{2-5} 
 & \multicolumn{4}{c}{$\Delta \log \Delta C_{it}$} \\ 
\\[-1.8ex] & (1) & (2) & (3) & (4)\\ 
\hline \\[-1.8ex] 
 lag(masks for employees only, 14) & $-$0.100$^{***}$ & $-$0.100$^{***}$ & $-$0.114$^{***}$ & $-$0.118$^{***}$ \\ 
  & (0.036) & (0.035) & (0.030) & (0.028) \\ 
  lag(masks in public spaces, 14) & $-$0.076$^{*}$ & $-$0.073$^{*}$ & $-$0.108$^{***}$ & $-$0.111$^{***}$ \\ 
  & (0.040) & (0.041) & (0.039) & (0.040) \\ 
  lag(closed K-12 schools, 14) & $-$0.153$^{***}$ & $-$0.153$^{***}$ & $-$0.047 & $-$0.049 \\ 
  & (0.054) & (0.054) & (0.066) & (0.065) \\ 
  lag(stay at home, 14) & $-$0.085 & $-$0.085 & $-$0.090$^{*}$ & $-$0.087 \\ 
  & (0.052) & (0.054) & (0.051) & (0.053) \\ 
  lag(business closure policies, 14) & $-$0.047 &  & 0.006 &  \\ 
  & (0.057) &  & (0.054) &  \\ 
  lag(closed movie theaters, 14) &  & 0.017 &  & 0.022 \\ 
  &  & (0.062) &  & (0.066) \\ 
  lag(closed restaurants, 14) &  & $-$0.054 &  & 0.002 \\ 
  &  & (0.054) &  & (0.058) \\ 
  lag(closed non-essent bus, 14) &  & $-$0.008 &  & $-$0.016 \\ 
  &  & (0.038) &  & (0.032) \\ 
  lag($\Delta \log \Delta C_{it}$, 14) & 0.057$^{**}$ & 0.059$^{**}$ & 0.058$^{**}$ & 0.058$^{**}$ \\ 
  & (0.023) & (0.024) & (0.026) & (0.026) \\ 
  lag($\log \Delta C_{it}$, 14) & $-$0.156$^{***}$ & $-$0.157$^{***}$ & $-$0.120$^{***}$ & $-$0.120$^{***}$ \\ 
  & (0.015) & (0.015) & (0.015) & (0.015) \\ 
  lag($\Delta \log \Delta C_{it}$.national, 14) &  &  & $-$0.125$^{***}$ & $-$0.126$^{***}$ \\ 
  &  &  & (0.039) & (0.044) \\ 
  lag($\log \Delta C_{it}$.national, 14) &  &  & $-$0.184$^{***}$ & $-$0.185$^{***}$ \\ 
  &  &  & (0.032) & (0.032) \\ 
  $\Delta \log T_{it}$ & 0.155$^{***}$ & 0.155$^{***}$ & 0.153$^{***}$ & 0.154$^{***}$ \\ 
  & (0.043) & (0.043) & (0.041) & (0.041) \\ 
 \hline \\[-1.8ex] 
state variables & Yes & Yes & Yes & Yes \\ 
Month $\times$ state variables & Yes & Yes & Yes & Yes \\ 
\hline \\[-1.8ex] 
$\sum_j \mathrm{Policy}_j$ & -0.461$^{***}$ & -0.456$^{***}$ & -0.353$^{**}$ & -0.357$^{**}$ \\ 
 & (0.134) & (0.135) & (0.139) & (0.142) \\ 
Observations & 3,651 & 3,651 & 3,651 & 3,651 \\ 
R$^{2}$ & 0.765 & 0.765 & 0.773 & 0.773 \\ 
Adjusted R$^{2}$ & 0.762 & 0.762 & 0.771 & 0.771 \\ 
\hline 
\hline \\[-1.8ex] 
\textit{Note:}  & \multicolumn{4}{r}{$^{*}$p$<$0.1; $^{**}$p$<$0.05; $^{***}$p$<$0.01} \\ 
\end{tabular} 

%% file: tables_and_figures/cases-maskall-long-aug-piy.tex
\begin{tabular}{@{\extracolsep{1pt}}lcccc} 
\\[-1.8ex]\hline 
\hline \\[-1.8ex] 
 & \multicolumn{4}{c}{\textit{Dependent variable:}} \\ 
\cline{2-5} 
 & \multicolumn{4}{c}{$\Delta \log \Delta C_{it}$} \\ 
\\[-1.8ex] & (1) & (2) & (3) & (4)\\ 
\hline \\[-1.8ex] 
 lag(masks for employees only, 14) & $-$0.060$^{**}$ & $-$0.057$^{**}$ & $-$0.062$^{***}$ & $-$0.058$^{**}$ \\ 
  & (0.023) & (0.025) & (0.021) & (0.023) \\ 
  lag(masks in public spaces, 14) & $-$0.121$^{***}$ & $-$0.110$^{***}$ & $-$0.124$^{***}$ & $-$0.111$^{***}$ \\ 
  & (0.029) & (0.033) & (0.029) & (0.033) \\ 
  lag(closed K-12 schools, 14) & $-$0.161$^{***}$ & $-$0.164$^{***}$ & $-$0.083$^{*}$ & $-$0.085$^{*}$ \\ 
  & (0.048) & (0.048) & (0.048) & (0.050) \\ 
  lag(stay at home, 14) & $-$0.059 & $-$0.067$^{*}$ & $-$0.056 & $-$0.066$^{*}$ \\ 
  & (0.038) & (0.038) & (0.038) & (0.039) \\ 
  lag(business closure policies, 14) & $-$0.107$^{**}$ &  & $-$0.077$^{*}$ &  \\ 
  & (0.045) &  & (0.046) &  \\ 
  lag(closed movie theaters, 14) &  & $-$0.096$^{***}$ &  & $-$0.094$^{***}$ \\ 
  &  & (0.028) &  & (0.028) \\ 
  lag(closed restaurants, 14) &  & 0.021 &  & 0.039 \\ 
  &  & (0.048) &  & (0.050) \\ 
  lag(closed non-essent bus, 14) &  & $-$0.020 &  & $-$0.007 \\ 
  &  & (0.039) &  & (0.038) \\ 
  lag($\Delta \log \Delta C_{it}$, 14) & 0.074$^{***}$ & 0.072$^{***}$ & 0.070$^{***}$ & 0.069$^{***}$ \\ 
  & (0.021) & (0.021) & (0.024) & (0.024) \\ 
  lag($\log \Delta C_{it}$, 14) & $-$0.135$^{***}$ & $-$0.136$^{***}$ & $-$0.111$^{***}$ & $-$0.113$^{***}$ \\ 
  & (0.010) & (0.011) & (0.011) & (0.012) \\ 
  lag($\Delta \log \Delta C_{it}$.national, 14) &  &  & $-$0.086$^{**}$ & $-$0.096$^{**}$ \\ 
  &  &  & (0.037) & (0.038) \\ 
  lag($\log \Delta C_{it}$.national, 14) &  &  & $-$0.143$^{***}$ & $-$0.147$^{***}$ \\ 
  &  &  & (0.025) & (0.025) \\ 
  $\Delta \log T_{it}$ & 0.140$^{***}$ & 0.140$^{***}$ & 0.138$^{***}$ & 0.137$^{***}$ \\ 
  & (0.042) & (0.042) & (0.040) & (0.040) \\ 
 \hline \\[-1.8ex] 
state variables & Yes & Yes & Yes & Yes \\ 
Month $\times$ state variables & Yes & Yes & Yes & Yes \\ 
\hline \\[-1.8ex] 
$\sum_j \mathrm{Policy}_j$ & -0.508$^{***}$ & -0.492$^{***}$ & -0.402$^{***}$ & -0.384$^{***}$ \\ 
 & (0.077) & (0.085) & (0.080) & (0.091) \\ 
Observations & 6,733 & 6,733 & 6,733 & 6,733 \\ 
R$^{2}$ & 0.676 & 0.677 & 0.682 & 0.683 \\ 
Adjusted R$^{2}$ & 0.673 & 0.674 & 0.679 & 0.681 \\ 
\hline 
\hline \\[-1.8ex] 
\textit{Note:}  & \multicolumn{4}{r}{$^{*}$p$<$0.1; $^{**}$p$<$0.05; $^{***}$p$<$0.01} \\ 
\end{tabular} 

%% file: tables_and_figures/cases-maskall-long-july-piy.tex
\begin{tabular}{@{\extracolsep{1pt}}lcccc} 
\\[-1.8ex]\hline 
\hline \\[-1.8ex] 
 & \multicolumn{4}{c}{\textit{Dependent variable:}} \\ 
\cline{2-5} 
 & \multicolumn{4}{c}{$\Delta \log \Delta C_{it}$} \\ 
\\[-1.8ex] & (1) & (2) & (3) & (4)\\ 
\hline \\[-1.8ex] 
 lag(masks for employees only, 14) & $-$0.064$^{**}$ & $-$0.056$^{*}$ & $-$0.067$^{***}$ & $-$0.062$^{**}$ \\ 
  & (0.028) & (0.030) & (0.026) & (0.029) \\ 
  lag(masks in public spaces, 14) & $-$0.101$^{***}$ & $-$0.097$^{**}$ & $-$0.126$^{***}$ & $-$0.124$^{***}$ \\ 
  & (0.037) & (0.042) & (0.039) & (0.044) \\ 
  lag(closed K-12 schools, 14) & $-$0.160$^{***}$ & $-$0.161$^{***}$ & $-$0.057 & $-$0.059 \\ 
  & (0.055) & (0.056) & (0.059) & (0.061) \\ 
  lag(stay at home, 14) & $-$0.064 & $-$0.073$^{*}$ & $-$0.065 & $-$0.075$^{*}$ \\ 
  & (0.040) & (0.041) & (0.040) & (0.041) \\ 
  lag(business closure policies, 14) & $-$0.089$^{*}$ &  & $-$0.043 &  \\ 
  & (0.048) &  & (0.052) &  \\ 
  lag(closed movie theaters, 14) &  & $-$0.099$^{**}$ &  & $-$0.086$^{**}$ \\ 
  &  & (0.040) &  & (0.041) \\ 
  lag(closed restaurants, 14) &  & 0.021 &  & 0.048 \\ 
  &  & (0.054) &  & (0.056) \\ 
  lag(closed non-essent bus, 14) &  & $-$0.011 &  & $-$0.008 \\ 
  &  & (0.040) &  & (0.040) \\ 
  lag($\Delta \log \Delta C_{it}$, 14) & 0.070$^{***}$ & 0.068$^{***}$ & 0.067$^{***}$ & 0.065$^{***}$ \\ 
  & (0.022) & (0.022) & (0.024) & (0.024) \\ 
  lag($\log \Delta C_{it}$, 14) & $-$0.143$^{***}$ & $-$0.143$^{***}$ & $-$0.115$^{***}$ & $-$0.115$^{***}$ \\ 
  & (0.012) & (0.013) & (0.013) & (0.014) \\ 
  lag($\Delta \log \Delta C_{it}$.national, 14) &  &  & $-$0.114$^{***}$ & $-$0.124$^{***}$ \\ 
  &  &  & (0.039) & (0.040) \\ 
  lag($\log \Delta C_{it}$.national, 14) &  &  & $-$0.178$^{***}$ & $-$0.180$^{***}$ \\ 
  &  &  & (0.031) & (0.033) \\ 
  $\Delta \log T_{it}$ & 0.138$^{***}$ & 0.137$^{***}$ & 0.135$^{***}$ & 0.134$^{***}$ \\ 
  & (0.039) & (0.038) & (0.037) & (0.037) \\ 
 \hline \\[-1.8ex] 
state variables & Yes & Yes & Yes & Yes \\ 
Month $\times$ state variables & Yes & Yes & Yes & Yes \\ 
\hline \\[-1.8ex] 
$\sum_j \mathrm{Policy}_j$ & -0.478$^{***}$ & -0.477$^{***}$ & -0.358$^{***}$ & -0.366$^{***}$ \\ 
 & (0.086) & (0.098) & (0.095) & (0.107) \\ 
Observations & 5,152 & 5,152 & 5,152 & 5,152 \\ 
R$^{2}$ & 0.692 & 0.693 & 0.699 & 0.700 \\ 
Adjusted R$^{2}$ & 0.689 & 0.690 & 0.697 & 0.698 \\ 
\hline 
\hline \\[-1.8ex] 
\textit{Note:}  & \multicolumn{4}{r}{$^{*}$p$<$0.1; $^{**}$p$<$0.05; $^{***}$p$<$0.01} \\ 
\end{tabular} 

%% file: tables_and_figures/cases-maskall-long-sept-piy.tex
\begin{tabular}{@{\extracolsep{1pt}}lcccc} 
\\[-1.8ex]\hline 
\hline \\[-1.8ex] 
 & \multicolumn{4}{c}{\textit{Dependent variable:}} \\ 
\cline{2-5} 
 & \multicolumn{4}{c}{$\Delta \log \Delta C_{it}$} \\ 
\\[-1.8ex] & (1) & (2) & (3) & (4)\\ 
\hline \\[-1.8ex] 
 lag(masks for employees only, 14) & $-$0.057$^{***}$ & $-$0.056$^{***}$ & $-$0.058$^{***}$ & $-$0.056$^{***}$ \\ 
  & (0.021) & (0.022) & (0.018) & (0.020) \\ 
  lag(masks in public spaces, 14) & $-$0.108$^{***}$ & $-$0.101$^{***}$ & $-$0.109$^{***}$ & $-$0.100$^{***}$ \\ 
  & (0.026) & (0.029) & (0.026) & (0.028) \\ 
  lag(closed K-12 schools, 14) & $-$0.149$^{***}$ & $-$0.152$^{***}$ & $-$0.072$^{*}$ & $-$0.074$^{*}$ \\ 
  & (0.044) & (0.044) & (0.040) & (0.042) \\ 
  lag(stay at home, 14) & $-$0.061$^{*}$ & $-$0.067$^{*}$ & $-$0.055 & $-$0.063$^{*}$ \\ 
  & (0.036) & (0.036) & (0.036) & (0.037) \\ 
  lag(business closure policies, 14) & $-$0.115$^{***}$ &  & $-$0.080$^{*}$ &  \\ 
  & (0.042) &  & (0.043) &  \\ 
  lag(closed movie theaters, 14) &  & $-$0.084$^{***}$ &  & $-$0.079$^{***}$ \\ 
  &  & (0.024) &  & (0.024) \\ 
  lag(closed restaurants, 14) &  & 0.014 &  & 0.029 \\ 
  &  & (0.048) &  & (0.049) \\ 
  lag(closed non-essent bus, 14) &  & $-$0.027 &  & $-$0.008 \\ 
  &  & (0.039) &  & (0.038) \\ 
  lag($\Delta \log \Delta C_{it}$, 14) & 0.075$^{***}$ & 0.074$^{***}$ & 0.069$^{***}$ & 0.069$^{***}$ \\ 
  & (0.021) & (0.021) & (0.023) & (0.023) \\ 
  lag($\log \Delta C_{it}$, 14) & $-$0.129$^{***}$ & $-$0.131$^{***}$ & $-$0.105$^{***}$ & $-$0.107$^{***}$ \\ 
  & (0.010) & (0.010) & (0.010) & (0.010) \\ 
  lag($\Delta \log \Delta C_{it}$.national, 14) &  &  & $-$0.081$^{**}$ & $-$0.087$^{**}$ \\ 
  &  &  & (0.036) & (0.037) \\ 
  lag($\log \Delta C_{it}$.national, 14) &  &  & $-$0.149$^{***}$ & $-$0.153$^{***}$ \\ 
  &  &  & (0.023) & (0.023) \\ 
  $\Delta \log T_{it}$ & 0.140$^{***}$ & 0.139$^{***}$ & 0.137$^{***}$ & 0.136$^{***}$ \\ 
  & (0.040) & (0.039) & (0.038) & (0.038) \\ 
 \hline \\[-1.8ex] 
state variables & Yes & Yes & Yes & Yes \\ 
Month $\times$ state variables & Yes & Yes & Yes & Yes \\ 
\hline \\[-1.8ex] 
$\sum_j \mathrm{Policy}_j$ & -0.490$^{***}$ & -0.472$^{***}$ & -0.374$^{***}$ & -0.351$^{***}$ \\ 
 & (0.072) & (0.077) & (0.074) & (0.082) \\ 
Observations & 8,314 & 8,314 & 8,314 & 8,314 \\ 
R$^{2}$ & 0.667 & 0.668 & 0.674 & 0.675 \\ 
Adjusted R$^{2}$ & 0.665 & 0.666 & 0.672 & 0.673 \\ 
\hline 
\hline \\[-1.8ex] 
\textit{Note:}  & \multicolumn{4}{r}{$^{*}$p$<$0.1; $^{**}$p$<$0.05; $^{***}$p$<$0.01} \\ 
\end{tabular} 